# Atmospheric instability as a loss of a metastable equilibrium

Yuri Kornyushin

*Maître Jean Brunschvig Research Unit, Chalet Shalva,
Randogne, CH-3975*

**Abstract.** *General thermodynamic theory of metastable states is used in this short note to try to understand better atmospheric instabilities. It is shown that not only cooling of a cloud can lead to rain, but heating also, especially when there are charged water drops in a cloud (in this case we have rain with lightning). The influence of the global warming on weather is discussed.*

## 1. Introduction

Metastable state in Classical Mechanics is a very well known one and it is shown on Fig. 1, which presents a schematic dependence of mechanical energy E on some parameter X (generalized coordinate). Metastable state (1) exists in a shallow minimum of mechanical energy. It is separated from more stable state (3) by a relatively low barrier (2). As the barrier is relatively low, the equilibrium in position (1) can be easily unstable.

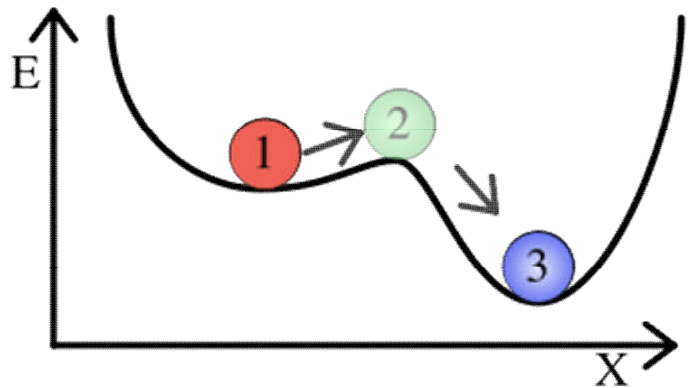

Fig. (1). Metastable state (1) in a shallow minimum of the energy in Classical Mechanics. It is separated from more stable state (3) by rather low barrier (2). Parameter X represents some generalized coordinate.

Metastable states in Classical Thermodynamics were very well known quite a long time ago (see, e.g. [1]). The specific feature of the metastable state in Classical Thermodynamics is that there exists some virtual critical point (temperature of the absolute instability), $T_i$. At this temperature the thermodynamic barrier, separating a metastable state from more stable one, does not exist. The matter is that the height of the barrier mentioned depends on temperature, $T$, and ceases to exist at $T = T_i$ (see, e.g., [2,3]). A number of metastable states were investigated theoretically and it was shown that these features are inherent for the metastable states in Classical Thermodynamics.

General thermodynamic theory of metastable states was first formulated in [2]. It was developed and published later in [3] also.

In general thermodynamic theory of metastable states entropy $S$ is used as a generalized thermodynamic coordinate [1,3]. It was shown that the thermodynamic barrier, separating metastable and more stable states is proportional to $(T - T_i)^{3/2}$ in a general case as a rule [1,3]. It could be different in some exotic cases [3]. So the stability of metastable states is described by general thermodynamic theory of metastable states. What happens after the loss of stability is not a subject of thermodynamic theory of metastable states.

## 2. Atmospheric instabilities

Atmospheric pressure, composition of the atmosphere, especially the level of humidity, and temperature, have a great impact on atmospheric stability. In case when the temperature $T$ is close to the virtual critical point $T_i$, a small increase in the entropy of a system can bring instability. A small heating brings a small increase in entropy, and temperature of a system becomes closer to $T_i$. So the instability occurs when the temperature of a system approaches $T_i$ [1,3].

When this collapse is accompanied by a fast formation of local high energetic movements, the energy of these movements comes from the internal energy of the atmosphere, thus leading to a fast decrease in temperature, according to the energy conservation law at almost adiabatic conditions.

Atmospheric instabilities took place usually in hot tropical seas. It looks like the humidity before the collapse is about saturation. So the data on temperature and pressure $P$ of the collapse could serve to estimate $T_i(P)$ and use these data for weather forecast.

One could imagine that when the average temperature of the atmosphere of the earth reaches certain value, the instability could occur on a global scale.

## 3. Overheated and overcooled phases

It is well known that metastable phases could be overheated and overcooled (see, e.g., [3]). For instance let us consider a cloud, consisting of air and two phases of water, small drops of water and a water vapor. Let us assume that the humidity is more than 100% relative to the flat surface of water. It is possible, because the water drops are small and the contribution of the surface energy to the saturating value of the pressure of a water vapor is essential. Then the water vapor phase could be regarded as an overcooled one. Cooling it further, the virtual critical point $T_i$ for this phase can be reached [3]. Then the overcooled water vapor phase will collapse into large water drops phase (rain). So, the cooling of a cloud may lead to some rain.

At the same time the water drops could be the overheated phase, if the water drops produce the partial water vapor pressure smaller than that, corresponding to the value of their radius. During further heating, $T_i$ for this phase could be reached; in this case the water drops phase could collapse into the water vapor phase, which could collapse into the large water drops phase (rain again, or shower). So, the heating of a cloud may lead to some rain also.

## 4. The role of the electric charge

Let us consider a drop of water of a radius $R$, charged with $N$ electrons (the electron charge is $e$). It is well known that calculating electrostatic energy of $N$ electrons, self-action should not be taken into account (see, e.g., [4]). The energy of the electrostatic field outside the drop is as follows [4]:

$$U = e^2 N(N-1)/2R. \qquad (1)$$

When the electric charge is situated on the surface of a drop, the electrostatic field in the volume of a drop is zero, and Eq. (1) represents the electrostatic energy of a charged drop. When the electric charge is distributed homogeneously in the volume of a drop, the energy of the electrostatic field inside the drop is [4] $e^2 N(N-1)/10\varepsilon R$, where $\varepsilon$ is the dielectric constant. Taking into account that for water $\varepsilon$ is about 80 at $T = 293$ K [5], one can see that the latest term is about 400 times smaller than the right-hand part of Eq. (1). This is a good reason to neglect the electrostatic field inside the drop volume while calculating the electrostatic energy of a charged drop.

As follows from the last two equations, the electrostatic energy of a drop, charged with one electron is zero. For $N = 2$ electrons we have

$$U_2 = e^2/R. \qquad (2)$$

The surface energy of the water drop, $4\pi\gamma R^2$ (here $\gamma$ is the surface energy of a unit area; for water $\gamma = 72.86$ erg/cm$^2$ at $T = 293$ K [6]) is equal to the electrostatic energy of a water drop, charged with two electrons, when $R = R_0 = (e^2/4\pi\gamma)^{1/3}$. For water the diameter of such a drop, $2R_0 = 1.263$ nm (this drop is a nano-object). It is well known that when the electrostatic energy of a water drop is larger than the surface energy, the drop disintegrates in two or more smaller drops [7]. So it is not reasonable to consider highly charged water drops.

Anyway, it is possible to include the electrostatic energy into effective surface energy $\gamma_e$,

$$\gamma_e = \gamma + e^2 N(N-1)/8\pi R^3. \qquad (3)$$

The electrostatic energy is inversely proportional to the radius of a drop. When the drop size decreases, the electrostatic energy increases. That is the charged drop is more difficult to evaporate than the neutral one. So the charge influences thermodynamic properties of a water drop. When charged drops collapse into the water vapor phase, discharge (lightning) accompanies this process.

## 5. Global warming and weather

Lately the weather forecast does not look reliable. It is difficult to understand the reasons for very frosty snowy winters and often rather cool summers in Western Europe. In the middle of January 2012 the author experienced rather a cool summer in Australia and New Zealand. It was snowing once in Melbourne in the middle of the summer 2012 (in the middle of January). The author assumes that these phenomena are some of the consequences of the global warming.

Global warming leads to the increase in the temperature, averaged over one cycle of time (one year) and over all the area of the surface of the Globe. It looks like the atmosphere is rather sensitive to a rather small increase in the average temperature. Indeed, average temperature of the atmosphere is about 300 K. Increase in the temperature by 1 K (increase of about 0.3 percent only) looks like rather an essential one. This gives some ground for assuming that the global atmosphere is in some metastable state and its averaged temperature is in the vicinity of a virtual critical point (the temperature of the absolute instability) [3]. It is reasonable to assume that in such a condition the slight increase in the temperature leads to the essential increase in the ventilation. Increasing ventilation leads to the decrease in the inhomogeneity of the temperature of the atmosphere. That is the colder areas, like around the poles, become warmer, and the warmer areas, like it is in Western Europe, Australia and New Zealand, become cooler.

Further increase in the average temperature of the atmosphere brings the atmosphere closer to the virtual critical point with more and more fluctuations [3]. When the average temperature becomes close enough to the virtual critical point, the metastable phase collapses into some more stable phase. What does it mean? The author assumes that it means rather a turbulent (stormy) phase. This phase possesses much more kinetic energy than the original one. According to the energy conservation law the average temperature of a new phase should be smaller than that of the original phase, may be like in the glacial epoch. Calculation shows that when some fraction of the kinetic energy of molecules of the atmosphere gives rise to the kinetic energy of aerodynamic movements the decrease in temperature $\delta T$ (in Kelvin degrees) could be approximated as $v^2/20000$ (here $v$ is the speed of the aerodynamic movement in the kilometer per hour units). That is at $v = 200$ kilometer per hour $\delta T = 2$ K, which is rather an essential quantity. The kinetic energy of the turbulent phase of the atmosphere will decay, leading to the increase in the average temperature of the atmosphere. Thus the cycle of the regarded transformations will be fulfilled.

The author could not tell anything about the duration of the described cycle yet.

## 6. Discussion

In this short note the author wanted to draw attention of the Atmospheric Thermodynamics research community and students to the option of regarding an atmospheric instability as a collapse of a local/global metastable state of the atmosphere. It is shown also that not only cooling of a cloud can bring rain, but heating also. Especially when there are charged water drops in a cloud. In this case a rain is accompanied by a lightning. Assuming that the atmosphere of the Globe is now in some metastable state close to the virtual critical point (the temperature of the absolute instability), the author infers a hypothetic cyclic in time behavior of the state of the atmosphere and evaluate the decrease in the temperature of the atmosphere, resulting from collapse of the original phase into more stable turbulent one.